\documentclass[12pt]{iopart}

\usepackage{color}

\usepackage{graphicx}
\usepackage{caption}

\begin{document}

\title{Phases of granular matter}

\author{Charles Radin$^1$ and Harry L. Swinney$^{2}$}

\address{$^{1}$Department of Mathematics, University of Texas at Austin, Austin, Texas 78712 USA}

\address{$^{2}$Department of Physics and Center for Nonlinear Dynamics, University of Texas at Austin, Austin, Texas 78712 USA}
\ead{radin@math.utexas.edu and swinney@chaos.utexas.edu}
\vspace{10pt}
%\begin{indented}
%\item[]24 August 2018
%\end{indented}
\begin{abstract}
An understanding of homogeneous  nucleation of crystalline structure from a disordered medium such as a liquid remains an important unsolved problem in condensed matter physics. Guided by the results from a number of experiments on granular and colloidal systems in the past two decades, including in particular observations of homogeneous nucleation in colloidal and granular systems, we suggest an alternative to the statistical mechanics approach to static granular matter initiated by Edwards and Oakeshott in 1989. 

\end{abstract}

\section{Introduction}

An important objective for a theoretical model of static, bulk, granular matter is to understand the unusual mechanical properties of the material. For instance, how should one understand why a very heavy vehicle can  stand on, or even move slowly over, a deep bed of dry sand?  This question was raised many years ago for materials composed of many molecules, where  heavy weights can be supported by ice but not by water; matter made of many $H_2O$  molecules may or may not support a heavy weight. Equilibrium statistical mechanics used the notions of fluid and solid phases to sharpen the question, and centered on the relevance  of temperature and pressure.

In 1989 Edwards and Oakeshott [1] developed a theoretical approach to the understanding of granular matter that was inspired by earlier modeling by Flory (see [2]), de Gennes (see [3]), and Edwards (see Doi and Edwards [4]) and others, who adapted the ensemble formalism of equilibrium statistical mechanics to model `soft' {\it nonequilibrium} matter such as a polymer chain of many segments in solution. One simplification in the adaptation for polymers is to downplay or ignore the need to justify the appropriateness of using ensemble averaging, which in statistical mechanics is based on the dynamics of the particles. 
Edwards and Oakeshott went further and applied the approach to {static} granular matter such as bulk sand, using microcanonical and canonical ensembles based on packings of nonoverlapping {\it stationary} spheres, mechanically stable under gravity and the contact forces among themselves and their container.

%a volume function that replaces the Hamiltonian. 
{The solidity of sand, a disordered assembly of many small grains, is reminiscent of window glass, a disordered collection of silica molecules whose bulk solidity is also poorly understood. These two old unsolved problems were conflated in 1998 [5] by a jamming paradigm which proposed  a common framework for properties of these and other materials, including colloids. A recent comprehensive review by Baule et al.\ [6] 
focuses on the statistical model of granular matter introduced by Edwards and Oakeshott and then generalized. We also note two papers of Kurchan and coworkers (see [7-8]) where the connection between the Edwards' model and glass theory is clearly described.} 
% \HS{PARENTHESES REMOVED}

In this paper we concentrate on {\it static} granular materials subjected to cyclic shear, where after each successive shear cycle it is possible to measure the position of the particles in the static system to high accuracy. Thus after the system is slightly perturbed by a small amplitude slow shear cycle [9], the change in position of each particle and the emergence of crystalline order can be detected. This situation contrasts with a glass where measurements of the positions of the individual silica molecules are not possible. Granular systems also contrast with collodial systems where precision measurement of the positions of individual colloidal particles is indeed possible, but the particles are not hard {[10]}, which complicates the determination of a well defined size. We discuss this and other differences between colloidal and granular systems in Section 3.2. 

{We use recent experiments on sheared granular systems undergoing cyclic vibration or cyclic shear of small amplitude for a wide range of frequencies to motivate a different approach to describe the solidity of static sand. We do not attempt a full model of the material throughout the {cyclic} motion. Our main interest is in the comparison of the state of the granular material at a typical high frequency, a typical low frequency, and zero frequency, and in particular on the dependence on the speed with which the frequency in reduced in between these regimes. We will discuss such a dynamical approach in connection with spontaneous homogeneous nucleation in granular matter.}

\section{Random close packing}

%\HS{(I agree, the Kepler concjecture is not relevant to our story; it is no longer mentioned.)} 

The configuration of grains in physical granular matter is generally disordered but difficult to characterize in detail. In 1960 G.D. Scott [11] conducted an experiment on ball bearings poured repeatedly into a container, and he found that the volume fraction $\phi$ occupied by the spheres ranged from about 0.60 up to a maximum of 0.64, which he conjectured to be a fundamental upper limit, now called the random close packed (RCP) volume fraction, $\phi_{RCP}$. The RCP volume fraction is much smaller than the maximal volume fraction occupied by a collection of identical spheres, which is the volume fraction of hexagonal close packed (HCP) or face centered cubic (FCC) crystalline arrays, $\phi=\pi/\sqrt{18}\approx 0.74$ [12].  

On the theoretical side, in 1959 J.D. Bernal [13] had already conjectured that ``there is an absolute impossibility of forming a homogeneous assembly of points of volume intermediate between those of long-range order and closest packed disorder''. There is an unstated probabilistic or entropy qualification, so to be more precise the claim is
that, in the sense of the Law of Large Numbers, 
there is a volume fraction, now called RCP, such that `most' homogeneous configurations of ball bearings  with fixed volume fraction below RCP are disordered and most homogeneous configurations with fixed volume fraction above RCP have long range order. 

In the half century following Bernal and Scott's introduction of the RCP concept of a fundamental limit in the volume fraction, many experiments have added increasing support for a fundamental limit in $\phi$ (see, e.g., [14-18]). {The notion of RCP has also been adapted to fit within theories of other materials, and even mathematical studies of sphere packings not closely associated with any material; {there is a good history of the generalized concept of RCP in Section III.4 of [6].} As noted in the previous section our proposals are accessible to physical experiment only for granular matter so we limit our use of the term RCP to its original sense as a physical phenomenon associated with bulk granular matter subjected to certain motions.} 

A recent experiment by Rietz et al.\ (2018)\ [9] {used} a  granular system with an imposed oscillating horizontal shear of small amplitude and low frequency.  This experiment revealed a gradual increase in $\phi$ until a plateau at $\phi=0.645$ was reached after 20000 shear cycles, as illustrated in Fig.\,1(a). The plateau  persisted for 50000 shear cycles, and then the global volume fraction began to slowly increase as a polycrystalline state developed {through { homogeneous} nucleation}.

An earlier experiment by Huerta et al.\ [19] also imposed a small amplitude horizontal oscillation but at a much higher frequency.  
The Huerta et al. and Rietz et al. experiments can both be thought of as granular systems in contact with a heat bath, where the frequency of oscillation is analogous to temperature. {(We apply the terms `heat bath' and `temperature' to nonequilibrium granular systems only in a vague intuitive sense, so that we may suggest an analogy with their well-defined meanings for ordinary materials made of molecules.)} In this {intuitive} terminology Huerta et al.\ used a bath at high temperature, 50 Hz, while Rietz et al.\ used a bath at low temperature, $0.5$ Hz. (The oscillation frequency was chosen to be low so that inertial effects were small; 
the typical Reynolds number for the spheres in the phthalate ester solution was $\sim\
 1$.) The two experiments can {then} be interpreted as indicating a fluid/solid phase transition with decreasing temperature, with the RCP volume fraction representing the nucleation barrier at the transition: the crystalline clusters need to reach a critical size, beyond which macroscopic crystalline matter automatically grows, as will be discussed in the Section 4. {We note that for our purposes it is mainly the dynamical nucleation process that is in need of detailed modeling, which is why we have used intuitive terms in describing the material under shear instead of attempting a serious model of such material. We will continue to treat the sheared material intuitively but will give more detail on nucleation later. }

\begin{figure}[ht!]
\centering
\includegraphics[width=\textwidth]{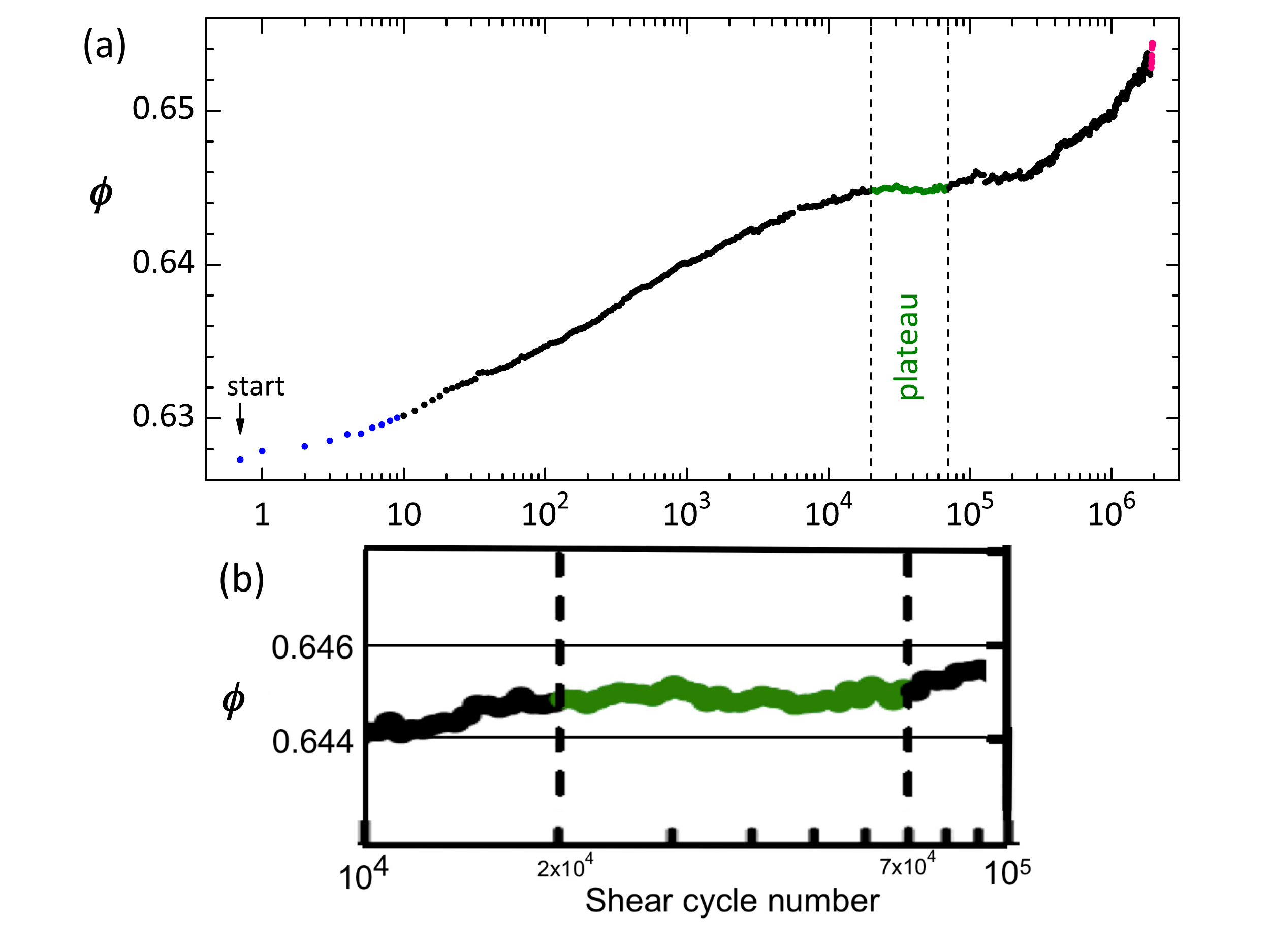}

\caption{(a) The robustness of the RCP limit is illustrated by these data of Rietz et al. [9] for a packing of 50000 glass spheres (3.00 mm diameter) in a $(105${~mm})$^3$ box with slow horizontal shear imposed by tilting two opposite sidewalls by only $\pm 0.6^{\circ}$ with an oscillation period of 2 s. The global volume fraction $\phi$ gradually increases with successive shear cycles until a well-defined plateau is reached at $\phi=0.645$. (b) An expanded graph of the plateau at $\phi=0.645$. The plateau persists for about 50000 shear cycles, and then the volume fraction begins to slowly increase as crystallites emerge in the interior of  the sample.
\label{fig1}}
\end{figure}

\section {A proposed alternative to the Edwards approach to static granular matter}

\noindent Edwards' fundamental idea of using an ensemble approach to model granular matter was very important as it suggested intuition based on materials in thermal equilibrium.  The Edwards approach has not proved to be helpful in understanding the random close packing phenomenon. A conjecture [20] that RCP might signify a sharp transition analogous to the freezing of fluids in thermal equilibrium was based on physical granular experiments showing crystalline clusters growing on side walls [21], not on calculations using an Edwards ensemble. There are now calculations suggesting that this interpretation of RCP is compatible with the Edwards model [22-23].
However, in subsections 4.1 and 4.3 it will be argued that the homogeneous nucleation observed in the experiment of  Rietz et al. [9] indicates that disordered static packings of grains should be considered as supercooled or glassy rather than in equilibrium as in the Edwards approach. {(We use the term `equilibrium' to refer to those states in a statistical granular theory, such as Edwards', which play the same role as the equilibrium states in the traditional statistical mechanics formalism of molecular materials.)}

We propose a theoretical approach for granular matter distinct from that of Edwards et al.\ but consistent with various laboratory observations.   An experiment by Huerta et al.\ [19] revealed buoyancy behavior like that given by Archimedes' Principle for an ordinary fluid in thermal equilibrium. As mentioned, many granular experiments have demonstrated the RCP limit, a well-defined upper bound on the volume fraction of homogeneous disordered granular matter, as illustrated by Fig.\ 1.  We interpret the experiments on buoyancy, the RCP limit, and the emergence of a crystalline state [9] as corresponding to granular matter at consecutively lower`temperatures'.
% , illustrating a fluid/solid freezing transition.
% We now describe in more detail several experiments that lead us to propose this alternative {approach} to the Edwards model of granular matter.}
% \CR{temperatures, akin to a fluid/solid freezing transition of ordinary matter, and now describe several physical experiments that lead us to propose this alternative to the Edwards treatment of static granular matter. The main difference in our approach is the emphasis on a dynamical nucleation process, rather than a statistical ensemble model, to understand the solidity of the material.}

\subsection{Archimedes' law of buoyancy }

Huerta et al.\ [19] studied a bidisperse collection of spheres, of diameters 3 and 4 mm, in a container with two opposite vertical side walls oscillating horizontally with small amplitude (1 mm) but high frequency (50 Hz) and acceleration (up to 10$g$). The high frequency oscillation maintained the granular medium in a fluidized state. Having the opposite walls oscillate with opposite phase eliminated convection. Low density objects pulled downward into the denser granular medium by a string, and dense test objects placed on top of the granular bed (away from the boundary) were observed to experience buoyant forces like those in an ordinary fluid in thermal equilibrium. Similar behavior was produced by shearing in Nichol et al. [24].

%\CR{I'll want to at least acknowledge Pouliquen et al}  \HS{(sure, fine)}. 

%\CR{The brief description of our experiment that used to be here somehow disappeared. I have resurrected it, but reference numbers may now be off.}

\subsection{Colloidal systems}

%\CR{I will rewrite this subsection in light of the analysis of the colloid experiment we discussed.} 

Extensive experiments have been conducted using concentrated suspensions of colloidal particles in baths in thermodynamic equilibrium.  Colloidal particles have been widely considered as hard spheres, but Poon et al.\ [25] and Royall et al.\ [10] explain how in reality colloidal particles are soft, as Fig.~2 illustrates.

\begin{figure}[ht!]
\centering
\includegraphics[width=125mm]{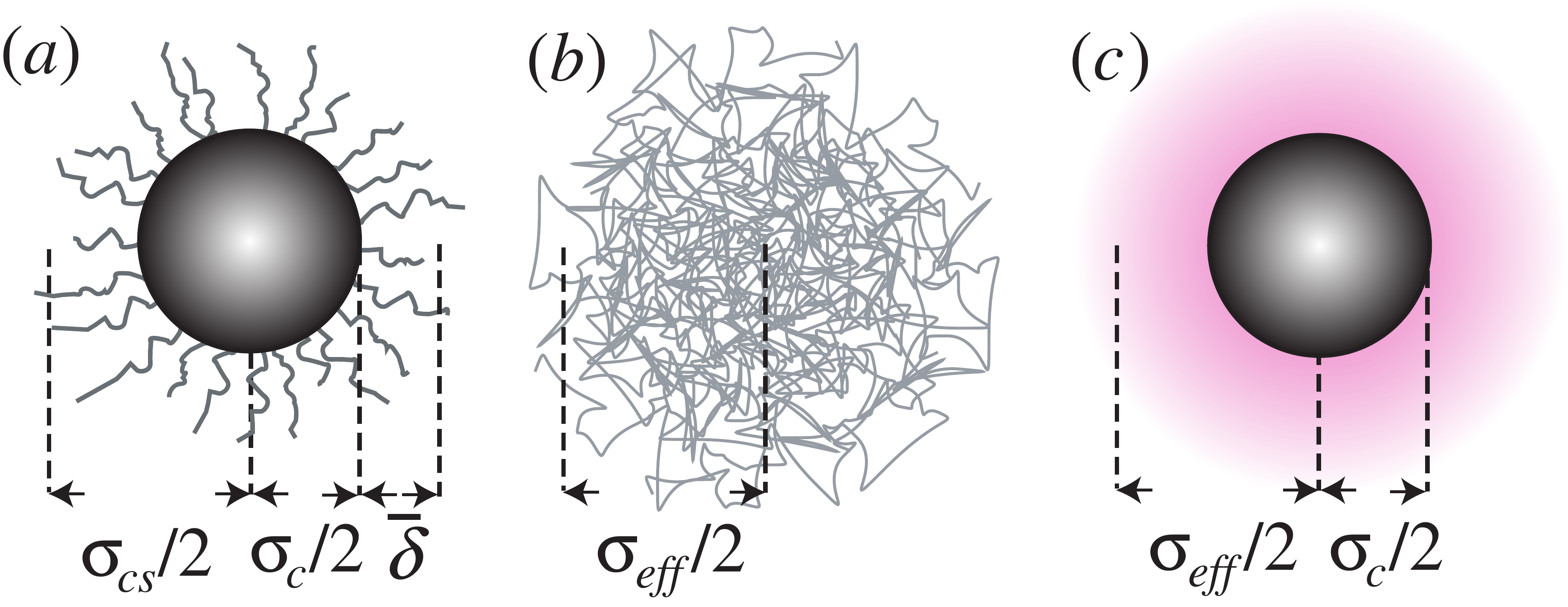}
\caption{Schematic representations of different models of ``hard-sphere" colloids (Fig. 1 in Royall et al.\ [10]):  (a) Sterically stabilized particle with a surface `hair'  layer of thickness $\bar{\delta}$ and core radius $\sigma_c /2$. (b) Microgel particle with an effective radius $\sigma_{eff} /2$, formed by a heavily cross-linked polymer. (c) Charged colloid with an electrical double-layer that gives an effective radius  $\sigma_{eff} /2$.
\label{fig2}}
\end{figure}
 
Another experiment on which we base our theoretical approach to granular  
matter was conducted by Rutgers et al.\ [26], who studied polystyrene spheres in water and used different methods to reduce electrostatic screening in order to mimic hard spheres more closely. There was still an uncertainty in the effective size of the particles, which was handled through an adjustable parameter. Figure~3 (from Fig.~2 of Rutgers et al. [26]) shows a sharp transition in density, which Rutgers et al.\ interpret as analogous to the fluid/crystal transition of the mathematical hard sphere model [27].

\begin{figure}[ht!]
\centering
\includegraphics[width=\textwidth]{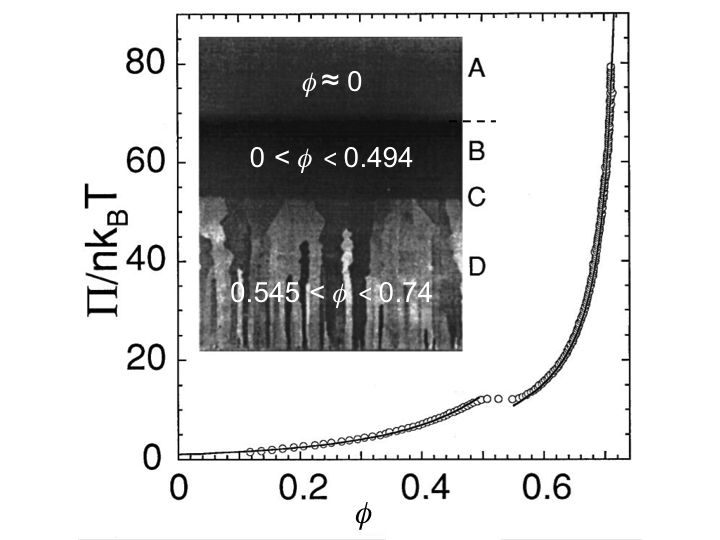}
\caption{The equation of state for the osmotic pressure $\Pi$ obtained by Rutgers et al. (Fig. 2 of [26]) for an isothermal system of polystyrene spheres (0.720 $\mu$m diameter) that was allowed to stabilize for three months before the measurements were made. The calculated equation of state, shown by a solid line, agrees well with the measurements of osmotic pressure. The inset shows 10 mm height of a sample: (A) supernatant fluid, (B) liquid phase (the dashed line indicates the faint boundary between the supernatant and the liquid phase), (C) sharp interface between the liquid phase and the phase with crystallites, and (D) phase with crystallites visible due to Bragg scattering of light.
\label{fig3}}
\end{figure}

The motion of colloidal particles is Brownian (diffusive), not ballistic as assumed in the hard sphere model, but such a simplification is not unusual for modeling soft materials; indeed, the particles in the Edwards model are stationary. A more accurate model would be based on Brownian motion, but such a model would only be appropriate for a
local analysis of individual particles, not of the bulk material, and would miss the bulk crystalline state shown in Rutgers et al.\ 
%\CR{(add a comma here)} \HS{(I think it's better without o comma, but I don't object to the addition of a comma.)} 
unless one takes into account gravity.
That experiment shows that gravity produces
particle assemblies with a significant vertical density 
gradient. Gravity and diffusive motion are both used in the model by Burdzy et al.\ [28], but they cannot analyze the region where Rutgers et al.\
see a discontinuous transition. Note that the hard sphere model employed by Rutgers et al. assumes ballistic motion but does not include gravity.

There are numerous differences between the behavior of the colloidal system in Rutgers et al.\ and the
granular system in Rietz et al., but there is also the remarkable similarity of the phase transition 
exhibited in the two. Below we will exploit common elements from Huerta et al.\ [19], Rietz et al. [9], and Rutgers et al.\ [26] in developing a heat bath picture of granular matter with `typical' configurations exhibiting a density gradient like those in the colloidal system of Fig.\ 3. 
% \CR{(We emphasize that colloids are not a focus of this paper.)}

\subsection{Dilatancy and the solidity of granular matter}

The homogeneous nucleation of crystalline structure in Rietz et al.\ indicates the creation of a `solid' state of granular matter. Crystallinity, observed for instance through X-ray diffraction for molecular systems, is a standard characteristic of many solids in thermodynamic equilibrium. (Quasicrystals are a notable exception.) {However, there are characteristics other than crystallinity} that might provide a clear distinction between solids and fluids, 
%A distinction between solids and fluids can also be based on macroscopic properties 
such as the viscosity or the elastic shear modulus [29].  We do not use such characteristics because there have yet to be experiments on mechanical properties of `crystalline' granular matter. However, there is reason to expect that such an approach should work. 

In 1885 Osborne Reynolds [30] introduced `dilatancy' to explain several experimental instances of the resistance of granular matter to shear. He imagined a {\it crystalline} arrangement of sand grains and argued that the arrangement would have to expand to be strained through a small angle, producing a strong force in reaction.

Dilatancy in a horizontally sheared granular system was demonstrated and measured by Nicolas et al.\ [21].  The system expanded when opposite side walls of the container were tilted slightly from the vertical position, thus decreasing the volume fraction, as shown in Fig.\ 4, where compaction and dilation occurs alternately through each successive cycle of shearing. Each successive maximum in the volume fraction was higher, as some particles were pulled down by gravity into space that opened up slightly in the dilated medium during the shearing process. In this experiment the volume fraction continued increasing past the RCP limit as  {\it inhomogeneous} nucleation of a crystal occurred at the cell walls (see Fig.\ 5 in [21]). 

\begin{figure}[ht!]
\centering
\includegraphics[width=125mm]
{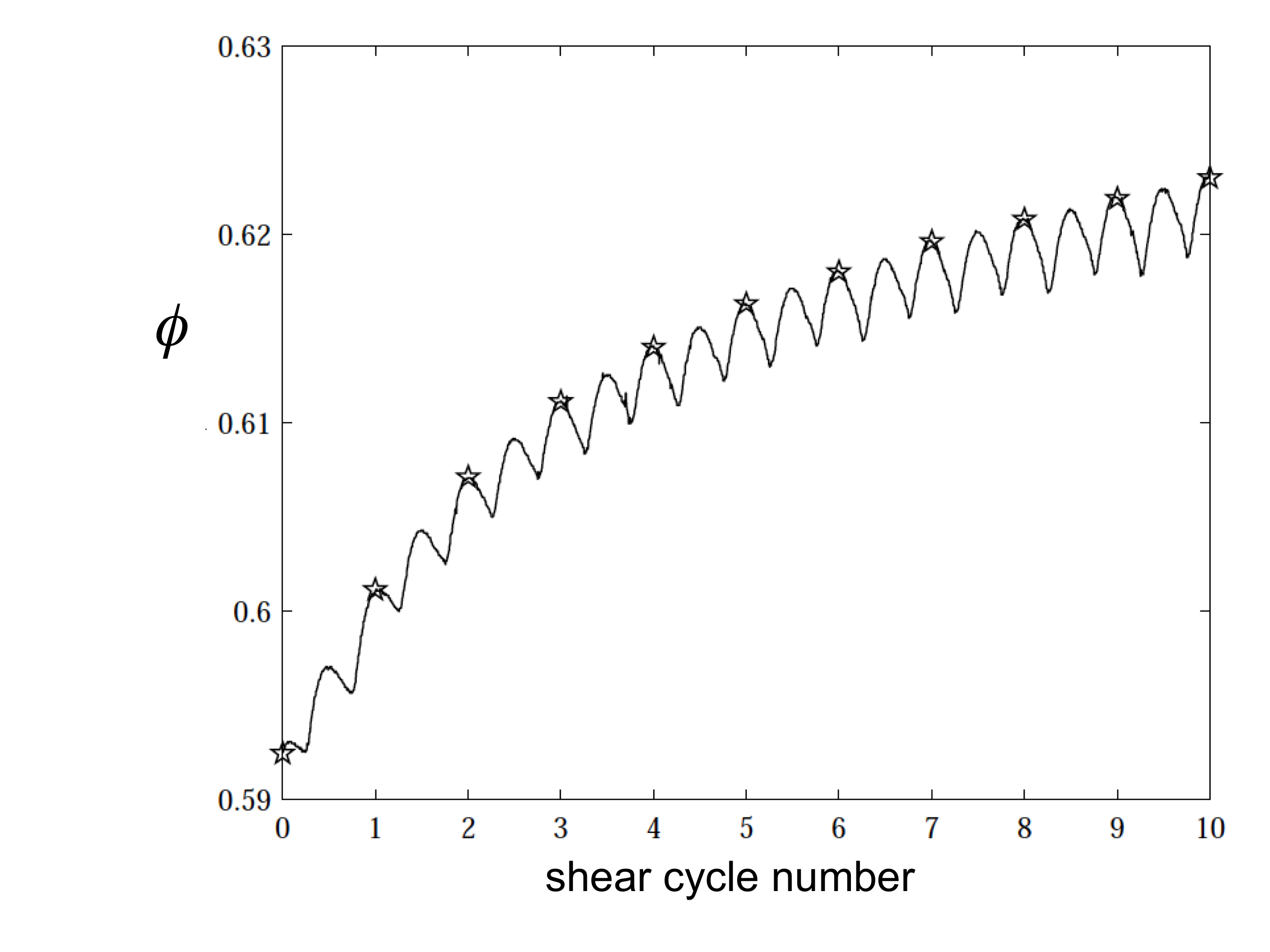}
\caption{The volume fraction evolution observed by Nicolas et al.\ [21] during successive cycles of shear of 41000 randomly distributed spheres in a parallelepiped box (Fig.\ 2 of [21]). Shear was imposed by oscillating the tilt angle of two opposite sidewalls by $\pm 5.4^\circ$.  The star symbols mark the volume fraction maxima in successive shear cycles as the sidewalls passed through the vertical position with the tilt angle increasing. 
\label{fig4}}
\end{figure}

\subsection{Dissipation in driven granular matter}

%\CR{I will rewrite this subsection.} 

For granular matter a comparison of experiments with simulations of molecular dynamics and granular continuum equations has revealed that dissipation through inelasticity (restitution coefficient less than unity) and sliding friction are both important in strongly driven systems, such as vertically oscillating granular layers. Extensive experiments and analyses on vertically oscillating granular layers has shown that when the downward acceleration of the container exceeds $g$, it is necessary to include both forms of dissipation (inelasticity and friction) to obtain agreement between observations and modeling [31-33]. It is not yet clear to what extent dissipation may be a necessary ingredient in the qualitative near equilibrium or equilibrium behavior we wish to understand in this paper.

A simple picture of the evolution of the granular matter in Rietz et al.\ is that the shearing mainly 
provides slow motion to the particles and breaks any contacts between
them, which allows mobility of the particles so they can occasionally
fall under gravity into holes that develop beneath them due to dilatancy (see Fig.\ 4). The particles in a colloid only diffuse, and the material as a whole stays somewhat localized (see Fig.\ 3 and [28]). In the granular experiment of Rietz et al.\ particles are mostly driven horizontally, and that, together with gravity, also leads to matter remaining near the floor. We will
use the similarity of the 
colloidal experiment of Rutgers et al.\ and the granular experiments of Huerta et al.\ and Nicolas et al.\ to fashion a common picture based on
a system of hard particles evolving under external energy input and gravity. We hope that such a picture will capture the essential qualitative features of the experiments, even ignoring dissipation. 
 
\section{Nucleation and glassy materials}

\subsection{Observations of homogeneous nucleation}
%\CR{I will rearrange this section, emphasizing experiment first: homo/inhomo nucl, then colloids, then CNT, then glass}

%\CR{The rest of this subsection will be put in section on nucleation.} 

Granular experiments with physical hard sphere packings have occasionally observed 
{\it inhomogeneous} nucleation at a bounding surface [21, 34-41], where two-dimensional layers easily form a triangular lattice that serves as the base for a three-dimension HCP/FCC cluster, like the oranges stacked in a market. In contrast, the recent experiment by Rietz et al.\ 
%with glass spheres (3.00 mm diameter) in an index-matched fluid 
revealed 
{\it homogeneous} nucleation, the spontaneous creation of crystalline clusters of `critical size' far from any bounding wall. An example of a growing crystallite in the granular experiment is shown in Fig.~5(a); crystallites of FCC, HCP, and mixed symmetries were observed to form and grow.  Similarly, an experiment on {\it colloidal} particles by of Gasser et al.\ [42] revealed homogeneous nucleation of crystallites, as illustrated by the example in Fig.~5(b). 
%\CR{Isn't the relevance of colloids unclear at this point? Perhaps we should split (b) from Figure 2?} 
The homogeneous nucleation of crystallites is a bulk property of the granular and colloidal systems, not a consequence of an interaction of the material with its boundaries, which is a surface phenomenon not necessarily indicative of any bulk property. 

\begin{figure}[ht!]
\centering
\includegraphics[width=125mm]{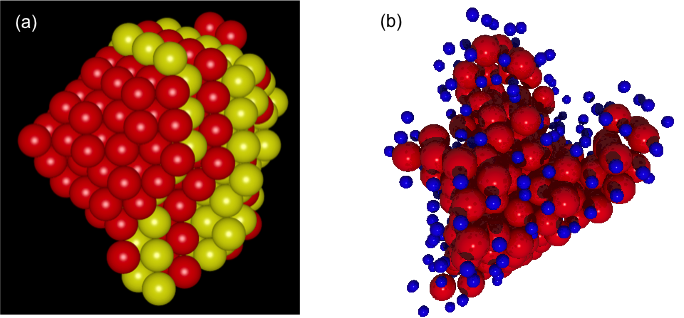}
\caption{(a) A crystallite with 600 spheres that nucleated in the interior of a sheared granular medium (unpublished figure from the experiment reported in Rietz et al.\ [9]). The red spheres have FCC local symmetry, and yellow spheres have HCP local symmetry.  (b)  A crystallite that Gasser et al.\ observed to nucleate following shear melting of a colloidal system with $\phi=0.47$ (Fig. 3C in [42]). The 206 red particles have crystalline order and are drawn to scale, while the neighboring blue particles, reduced in size for clarity, share at least one crystal-like bond. }
\label{fig5)}
\end{figure}

{Nucleation has been most thoroughly analyzed for {molecular} materials in thermal equilibrium{, although precise measurements of the molecular positions  in such materials are impractical} [43].} A static fluid in thermal equilibrium at fixed pressure and slowly decreasing temperature will freeze at a `freezing' temperature $T_f$ to a solid, typically a crystalline solid. With a slowly increasing temperature that solid will melt at a `melting' temperature $T_m$. With care one finds $T_f=T_m$, although a fluid can easily be driven with decreasing temperature to a supercooled (metastable, nonequilibrium) state at some temperature $T<T_f$, where $T$ is the container temperature.  Quasistatic freezing of a fluid to its equilibrium solid state generally occurs through nucleation, but unless care is taken to avoid it, the nucleation is inhomogeneous. If such nucleation sites are absent the fluid will supercool until it nucleates spontaneously, so-called homogeneous nucleation, at unpredictable {interior} sites. If a fluid is cooled sufficiently quickly, it will  fail to nucleate at all and will enter a nonequilibrium long-lived glassy state.

Homogeneous nucleation appears in fluids in thermal equilibrium as the temperature is slowly reduced below the freezing point. Within a thermodynamic phase, mechanical and other responses of matter in thermal equilibrium can be estimated by linear response of the equilibrium state.  A different approach, not necessarily part of a theory of bulk matter, is needed to model the nucleation of a new phase in response to a change of state  at a phase transition.

The observation in [9] of homogeneous nucleation in static granular  matter subjected to gentle cyclic shear suggests that one consider the static {disordered granular} material to be {analogous to molecular matter which is} either glassy or supercooled, but not in equilibrium.

\subsection{Nucleation theory}
The observations of nucleation in granular and colloidal matter could perhaps provide insight into nucleation more generally. The current state of the art in nucleation theory is a collection of models called Classical Nucleation Theory (CNT) [43], which has developed over many years to describe the spontaneous development of small crystalline clusters in a fluid.  In CNT there exists a `critical' size of a crystalline cluster which, once surpassed, will grow to macroscopic size.  Such a critical size has been observed in the recent experiments on nucleation in colloidal and granular systems, as Fig.\ 6 illustrates: in the colloidal system crystalline clusters with fewer than $\sim 100$ particles shrank while larger clusters grew (Gasser et al.\ [42]), and in the granular experiment  crystalline clusters with fewer than $\sim 10$ spheres shrank while larger clusters predominantly grew.\ 

While Classical Nucleation Theory provides a qualitative picture of nucleation that is consistent with observations, it is not a fundamental theory that can be tested quantitatively in physical experiments. Perhaps detailed measurements of nucleation in colloidal and granular systems will guide the development of a quantitative theory of nucleation that can be tested in experiments on both macroscopic and microscopic systems.

\begin{figure}[ht!]
\centering
\includegraphics[width=100mm]{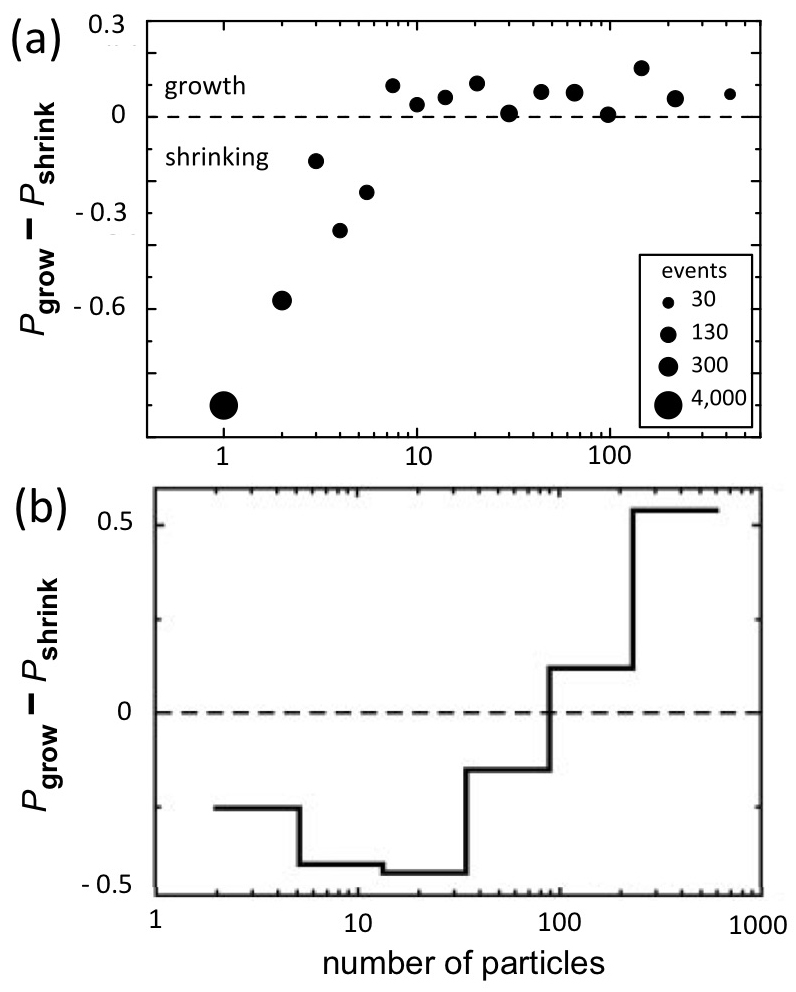}
\caption{ In both the granular and colloidal systems small crystalline clusters usually shrink while crystallites above a critical size grow, as demonstrated by these results: (a) in sheared granular matter (from Fig. 4(a) of Rietz et al.\ [9]), and (b) in a colloidal system with $\phi$=0.47 (Fig. 2 of Gasser et al.\ [42]). 
}
\label{fig6)}
\end{figure}

\subsection{Glassy states of granular matter}

Nucleation is significant because of the difficulty of a (disordered) fluid to freeze to an ordered (crystalline) solid. {Understanding the barriers to nucleation} is an essential part of the `glass problem'. One view of the granular experiment of Rietz et al.\ [9]  is that the experiment revealed an appropriate physical protocol to `equilibrate' supercooled granular matter to a crystalline state, smoothly but with an observable nucleation barrier, which turned out to explain {the physical phenomenon of granular} RCP. This suggests that granular matter under cyclic shear can provide an unusual setting to help understand some of the phenomena involved in glass formation. 
A similar claim has been made [42] for the nucleation of hard colloids, but precision for colloids is limited by the 3-6\% uncertainty in the determination of particle size, which is illustrated by the images in Fig.\ 2 (from [10]).

Any sufficiently large volume of gas such as the atmosphere, or even a typical small solid such as a rock conglomerate, 
is heterogeneous at the largest scale, and basic physical theory is  usually applied
first to homogeneous parts and then extended to the heterogeneous
material.  In particular, we expect typical granular matter, such as a sand pile, to be heterogeneous in density
(and composition). At least since the 1996 experiment by Nowak et al.\ [44], researchers  have
tried multi-axis shaking [34-36], cyclic shear [21, 37-38], continuous shear [24, 39-40] and sedimentation [41] to prepare homogeneous  
%\HS{(deleted: static)}  
granular matter to which
one could apply the Edwards theory with its ensembles of {\it static} spheres.
In contrast, when a granular material is fluidized as in [19] or [24] and the external vibration is quickly
removed, the static material can be considered  `glassy', with mechanical properties different
from, and harder to analyze than, those of the fluid{ized state}. The interpretation that
the static state is the result of such a quench derives from 
the	difference of its disordered state from the granular matter at the end of the experiment of Rietz et al., at low but still positive {shear rate}. The result of quenching from such a crystalline state would still be crystalline, and quite different from commonly observed disordered granular matter.

The above discussion on the glassy state of granular matter should be distinguished from those theories which associate a glassy state, or transition to a glassy state, in analyses of hard sphere models without the ingredient of gravity; see for example [42].

\section{Conclusions}

We have sketched a way to understand the properties of granular matter composed of grains moving in response to a certain type of {external forcing,} explicitly drawing the approach from physical experiments that reveal relevant 
%\HS{(interesting changed to "relevant")} 
properties of granular and colloidal matter. The present view uses well-defined properties of granular and colloidal matter to explain how control parameters can turn granular matter from fluid-like to solid-like behavior through a sharp phase transition. Along the way this approach interprets several {physical} experiments on {random close packing (RCP)} that are outside the purview of the Edwards model, {as the latter is not structured to incorporate the dynamical process of nucleation that we utilize.}

With this perspective on granular matter it is {still} natural to seek an ensemble model such as that of Edwards, from which one could 
in principle compute material properties.
However, if the interest were ultimately in the properties of static granular matter, as in the solidity of a dry bed of sand as discussed in the Introduction, such a model would only be of use for (rare) systems with volume fraction above RCP, i.e.,\ above $\phi=0.645$.
In our approach  static systems with density below RCP are treated as glassy, and an ensemble model of  fluidized granular matter would be of little direct use.  That said, the experiment of Rietz et al.\ exhibits a simple physical protocol, namely, very small angle (0.6$^{\circ}$) and very low rate ($0.5$ Hz) of shear
perpendicular to gravity, which `equilibrates' any static initial state into a crystalline state; hence high density granular matter is now a natural target for study. Aside from the inherent interest in this crystalline state of physical granular matter, this protocol 
%\HS{(it-->this protocol) } 
would also be of potential use in investigating the general phenomenon of the nucleation of crystalline structure out of disorder. {In particular, we note that using this protocol experiments could easily be performed by gathering detailed physical data on the positions in time of granular grains as nucleation occurs, a circumstance not available in other materials and devoid of the many assumptions in current nucleation theories. And finally we argue that this approach, by treating static sand as a limit of cyclically sheared sand but with an emphasis on the speed with which the frequency is decreased, is a practical way to analyze the problem of the solidity of static sand, testable by physical experiment.}

\section{Acknowledgments}
%\HS{(ok?:: paragraph on Hohenberg has been moved to be before the standard acknowledgments.)}

One of the authors (Harry Swinney) acknowledges with pleasure the influence of Pierre Hohenberg, who is being remembered by the publication of this special memorial issue of the Journal of Statistical Physics.  In the period 1971-73 Pierre Hohenberg was a regular visitor in the physics department at New York University, where Harry was an assistant professor who was conducting light scattering studies of critical phenomena in fluids. Regular discussions with Pierre were particularly helpful in interpreting data for the decay rate of order parameter fluctuations near critical points. In the following decades there were many lively discussions where Pierre guided Harry in the interpretation of laboratory data in terms of statistical theory.

The authors of this paper acknowledge many lively and productive discussions with Matthias Schroeter, who initiated together with Frank Rietz and the authors the experiment at the University of Texas that motivated the present manuscript. The results presented in  Rietz et al.\ [6] were subsequently obtained at the Max Planck Institute for Dynamics and Self-Organization in Goettingen, where Schroeter and Rietz developed a much improved apparatus. 

The authors also thank Eric Weeks for helpful discussions about the experiments on nucleation in colloidal systems. 

Radin acknowledges the support of NSF grant DMS-1509088, and Swinney acknowledges the support of the Sid W. Richardson Foundation.\\

%\CR{I will put in the references here}

Bibliography\\

1 Edwards, S., Oakeshott, R.: Theory of powders,
Physica A (Amsterdam) {\bf 157}, 1080-1090 (1989)

2 Flory, P.: Statistical Mechanics of Chain Molecules, Hanser-Gardner,
Cincinnati, Ohio (1989)

3 de Gennes, P.G.: Soft matter,
Rev. Mod. Phys. {\bf 64}(3), 645-648 (1992)

4 Doi, M.,  Edwards, S.M.: The Theory of Polymer Dynamics,
International series of monographs on physics, Oxford University Press, England {\bf 73} (1986)

5 Liu, A., Nagel, S.: Jamming is not just cool any more,
Nature (396) 21-22 (1998)

6 Baule, A., Morone, F., Herrmann, H., Makse, H.: Edwards statistical
mechanics for jammed granular matter, Rev. Mod. Phys. {\bf 90}(1),
015006 1-58 (2018)

7 Barrat, A., Kurchan, J., Loreto, V., Sellitto, M.: Edwards' measures for powders and glasses, Phys. Rev. Lett. {\bf 85} 5034-5037 (2000)

8 Charbonneau, P., Kurchan, J., Parisi, G.,, Urbani, P., Zamponi, F.: Glass and jamming transitions: from exact results to finite-dimensional descriptions, Annu. Rev. Condens. Matter Phys. {\bf 8} 265-288 (2117)

9 Rietz, F., Radin, C., Swinney H.L., Schroeter, M.: Nucleation in sheared
granular matter, Phys. Rev. Lett. {\bf 120}(5), 055701  (2018)

10 Royall, C., Poon W., Weeks, E.: In search of colloidal hard
spheres, Soft Matter {\bf 9}, 17-27 (2013)

11 Scott, G.D.: Packing of equal spheres, Nature (London) {\bf 188},
908-909 (1960)

12 The Kepler Conjecture, ed. Lagarias, J.: Springer-Verlag, New York (2011)

13 Bernal, J.D.: A geometrical approach to the structure of
liquids, Nature (London) {\bf 183}, 141-147 (1959)

14 Scott, G.D., Kilgour, D.M.: 
The density of random close packing of spheres,
J. Phys. D {\bf 2}, 863-866 (1969)

15 Finney, J.L.: Random packings and the structure of simple liquids. I. The geometry of
random close packing,         
Proc. Roy. Soc. A {\bf 319}, 479-493 (1970)

16 Berryman, J.G.:
Random close packing of hard spheres and disks, 
Phys. Rev. A {\bf 27}, 1053-1061 (1983)

17 Aste, T., Saadatfar, M., and Senden, T. J.: Geometrical structure of disordered sphere packings, Phys. Rev. E {\bf 71}(6), 061302 (2005)

%13 Kapfer, S.C., Mickel, W., Mecke, K., Schroeder-Turk, G.E.:Jammed spheres: Minkowski tensors reveal onset of local crystallinity,Phys. Rev. E {\bf 85}, 030301(R) (2012)

18 Baranau, V., Tallarek, U.: 
Random-close packing limits for monodisperse and polydisperse hard spheres,
Soft Matter 10,
3826-3841 (2014).

19 Huerta, D.A., Sosa, V., Vargas, M.C., Ruiz-Suarez, J.C.:
Archimedes' principle in fluidized granular systems, Phys. Rev. E {\bf 72}(3),
031307  (2005)

%9 Kepler, J.: Strena seu de nive sexangula (The six-cornered
%snowflake), http://www.thelatinlibrary.com/kepler/strena.html, (1611)

%10a Lubachevsky, B.D., Stillinger, F.H.: Geometric properties of
%random disk packings, J. Stat. Phys. {\bf 60}(5-6), 561-583 (1990)

20 Radin, C.: Random close packing of granular matter,
J. Stat. Phys. {\bf 131}(4), 567-573 (2008)

21 Nicolas, M., Duru, P., Pouliquen, O.: Compaction of a granular
material under cyclic shear, Eur. Phys. J. E {\bf 3}, 309–-314 (2000)

22 Aristoff, D., Radin, C.: Random close packing in a granular model,
J. Math. Phys. {\bf 51}(11), 113302 (2010)

23 Jin, Y., Makse, H.A.: A first-order phase transition defines the
random close packing of hard spheres, Physica A (Amsterdam) {\bf 389},
5362-5379 (2010)

24 Nichol, K., Zanin, A., Bastien, R., Wandersman, E.,
van Hecke, M.: Flow-induced agitations create a granular fluid,
Phys. Rev. Lett. {\bf 104}(7) 078302 (2010)

25 Poon, W., Weeks E., Royall, C.: On measuring colloidal volume
fractions, Soft Matter {\bf 8}, 21-30 (2012)

26 Rutgers, M.A., Dunsmuir, J.H., Xue, J.-Z., Russel, W.B., Chaikin,
P.M.: Measurement of the hard-sphere equation of state using screened
charged polystyrene colloids, Phys. Rev. B {\bf 53}(9), 5043-5046 (1996)

27 Loewen, H.: Fun with hard spheres, pages 295-331 in Statistical
physics and spatial statistics: the art of analyzing and modeling
spatial structures and pattern formation, ed. K.R. Mecke and
D. Stoyen, Lecture notes in physics No. 554, Springer-Verlag, Berlin
(2000).

28 Burdzy, K., Chen Z.-Q., Pal, S.: Archimedes' Principle for Brownian
Liquid, Ann. Appl. Probab. {\bf 21}, 2053–2074 (2011)

29 Anderson, P.W.: Basic Notions of Condensed Matter Physics, Benjamin/Cummings, Menlo Park, Chap.\ 1, Sec E  (1984)

30 Reynolds, O.: On the dilatancy of media composed of rigid particles
  in contact, Phil. Mag. Series 5 {\bf 20},
  469-481 (1885)

31 Bizon, C., Shattuck, M.D., Swift, J.B., McCormick, W.D.,
Swinney, H.L.: Patterns in 3D vertically oscillated granular layers:
simulation and experiment, Phys. Rev. Lett. {\bf 80}(1), 57-60 (1998)

32 Moon, S., Swift, J.B., Swinney, H.L.: Role of friction in pattern
formation in oscillated granular layers, Phys. Rev. E {\bf 69}(3),
031301  (2004)

33 Bougie, J.,  Kreft, J., Swift, J.B., Swinney, H.L.: Onset of patterns
in an oscillated granular layer: continuum and molecular dynamics
simulations, Phys. Rev. E {\bf 71}(2), 021301 (2005)

34 Francois, N., Saadatfar, M., Cruikshank, R., and Sheppard, A.: Geometrical frustration in amorphous and 
partially crystallized packings of spheres, Phys. Rev. Lett. {\bf 111}, 148001 (2013)

35 Hanifpour, M., Francois, N., Robins, V., Kingston, A., Vaez Allaei, S. M., and Saadatfar, M.:
Structural and mechanical features of the order-disorder transition in experimental hard-sphere packings,
Phys. Rev. E {\bf 91}, 06202 (2015)

36 Saadatfar, M., Takeuchi, H., Robins, V., Francois, N.: Pore configuration landscape of granular
crystallization, Nature Commun. {\bf 8}, 1-11, 15082 (2017) DOI: 10.1038/ncomms15082

%34 Bernal, B., Cherry, I., Knight, K.R.: Growth of crystals from random close packing, Nature {\bf 202}, 852-854 (1964)\

37 Mueggenburg, N. W.: Behavior of granular materials under cyclic shear, Phys. Rev. E {\bf 71}, 031301 (2005)

38   Panaitescu, A., Reddy, K. A., Kudrolli, A: Nucleation and crystal growth in sheared granular sphere
packings, Phys. Rev. Lett. {\bf 108}, 108001 (2012).

39 Tsai, J. C., Voth, G. A., Gollub, J. P.: Internal granular dynamics, shear-induced crystallization, and compaction steps,
Phys. Rev. Lett. {\bf  91}, 064301 (2003)

40 Daniels, K.E., Behringer, R.P.: Characterization of a freezing/melting transition in a vibrated and
sheared granular medium, J. Stat. Mech. P07018 (2006)

41 Schroeter, M., Goldman, D.I., Swinney, H.L.: Phys. Rev.
E {\bf 71}, 030301(R) (2005).

42 Gasser, U., Weeks, E.R., Schofield, A., Pusey P.N., 
Weitz, D.A.: Real-space imaging of nucleation and growth in colloidal
crystallization, Science {\bf 292}, 258-262 (2001)

43 Debenedetti, P.G.: Metastable Liquids, Princeton University Press,
Princeton (1996)

44 Nowak, E., Knight, J., Ben-Naim, E., Jaeger H., Nagel, S.:
Density fluctuations in vibrated granular materials, Phys. Rev. E {\bf
57}(2), 1971-1982 (1998)

%45 Kamien R., Liu, A.: Why is Random Close Packing Reproducible?,Phys. Rev. Lett. {\bf 99}(15), 155501  (2007)\\

\end{document}